\documentclass[letterpaper, 10pt, conference]{ieeeconf}      

\IEEEoverridecommandlockouts                             

\usepackage[utf8]{inputenc}
\usepackage{graphicx}
\usepackage[acronym]{glossaries}
\usepackage{float}
\usepackage{comment}
\usepackage{amssymb}
\usepackage{hyperref}
\usepackage[capitalise]{cleveref}
\usepackage{fancyhdr}

\title{\LARGE\bf Accounting for the Effects of Probabilistic Uncertainty \\ During Fast Charging of Lithium-ion Batteries} 

\author{Minsu Kim$^{1}$, Joachim Schaeffer$^{1,2}$, Marc D. Berliner$^{1}$, Berta Pedret Sagnier$^{1}$, \\ Rolf Findeisen$^{2}$, and Richard D. Braatz$^{1,*}$, \textit{IEEE Fellow}
\thanks{$^1$ Massachusetts Institute of Technology, Cambridge, MA, USA.}
\thanks{$^2$ Technical University of Darmstadt, Germany.}%
\thanks{$^*$ Corresponding Author, {\tt\small braatz@mit.edu}}%
}

\newacronym{pet}{PET}{porous electrode theory}
\newacronym{dfn}{DFN}{Doyle-Fuller-Newman}
\newacronym{p2d}{P2D}{pseudo-two-dimensional}
\newacronym{ecm}{ECM}{equivalent circuit model}
\newacronym{lib}{LIB}{lithium-ion batteries}
\newacronym{lfp}{LFP}{lithium-iron-phosphate}
\newacronym{uq}{UQ}{uncertainty quantification}
\newacronym{pce}{PCE}{polynomial chaos expansion}
\newacronym{apce}{aPCE}{arbitrary polynomial chaos expansion}
\newacronym{mc}{MC}{Monte Carlo}

\newacronym{nmc}{NMC}{nickel-manganese-cobalt}
\newacronym{nca}{NCA}{nickel-cobalt-aluminum}
\newacronym{lli}{LLI}{loss of lithium inventory}
\newacronym{lam}{LAM}{loss of active material}
\newacronym{spm}{SPM}{single-particle model}
\newacronym{kmc}{kMC}{kinetic Monte Carlo}
\newacronym{mpet}{MPET}{Multiphase PET}
\newacronym{sei}{SEI}{solid-electrolyte interface}
\newacronym{dae}{DAE}{differential-algebraic equation}
\newacronym{soc}{SOC}{state-of-charge}
\newacronym{soh}{SOH}{state-of-health}
\newacronym{doc}{DoC}{depth-of-charge}
\newacronym{dod}{DoD}{depth-of-discharge}
\newacronym{cccv}{CC-CV}{constant current--constant voltage}
\newacronym{ols}{OLS}{Ordinary least squares}
\newacronym{lar}{LAR}{Least angle regression}
\newacronym{ccctcv}{CC-CT-CV}{constant current--constant temperature--constant voltage}
\newacronym{ct}{CT}{constant temperature}
\newacronym{em}{EM}{electrochemical model}
\newacronym{cc}{CC}{constant current}
\newacronym{cv}{CV}{constant voltage}
\newacronym{ci}{CI}{confidence interval}
\newacronym{qoi}{QoI}{quantity of interest}

\fancypagestyle{FirstPage}{
\lfoot{Accepted for ACC 2024 \copyright2024 IEEE} 
}

\begin{document}

\maketitle
\thispagestyle{empty}
\pagestyle{empty}

\begin{abstract}

Batteries are nonlinear dynamical systems that can be modeled by Porous Electrode Theory models. The aim of optimal fast charging is to reduce the charging time while keeping battery degradation low. Most past studies assume that model parameters and ambient temperature are a fixed known value and that all PET model parameters are perfectly known. In real battery operation, however, the ambient temperature and the model parameters are uncertain. To ensure that operational constraints are satisfied at all times in the context of model-based optimal control, uncertainty quantification is required. Here, we analyze optimal fast charging for modest uncertainty in the ambient temperature and 23 model parameters. Uncertainty quantification of the battery model is carried out using non-intrusive polynomial chaos expansion and the results are verified with Monte Carlo simulations. The method is investigated for a constant current--constant voltage charging strategy for a battery for which the strategy is known to be standard for fast charging subject to operating below maximum current and charging constraints. Our results demonstrate that uncertainty in ambient temperature results in violations of constraints on the voltage and temperature. Our results identify a subset of key parameters that contribute to fast charging among the overall uncertain parameters. Additionally, it is shown that the constraints represented by voltage, temperature, and lithium-plating overpotential are violated due to uncertainties in the ambient temperature and parameters. The C-rate and charge constraints are then adjusted so that the probability of violating the degradation acceleration condition is below a pre-specified value. This approach demonstrates a computationally efficient approach for determining fast-charging protocols that take probabilistic uncertainties into account.
\end{abstract}

\thispagestyle{FirstPage}

\section{Introduction}
Intercalation-based electrochemical batteries such as \gls{lib}s are critical components for sustainable energy grids, transportation, and mobile devices. Among various battery materials, \gls{lib}s are the most widely used due to their characteristics of high energy density, high open-circuit voltage, low self-discharge characteristics, and long lifetime. Long charging times and reduced capacity due to battery deterioration, however, still remain challenges. As such, a goal of the battery industry is to develop technologies that enable fast charging while minimizing degradation.

Many battery models have been used to model the cycling behavior of \gls{lib}s. Each battery model can be categorized as being an \gls{ecm} or an \gls{em}. \gls{em}s describe internal phenomena such as temperature distribution, concentration distribution, and \gls{sei} growth during charging, which is challenging for an \gls{ecm}. The most widely used \gls{em} is the \gls{pet} model, also called the \gls{p2d} model \cite{NEWMANN1975PET}. Much research has been devoted to improving the modeling of degradation (e.g., \cite{Edge_OKane_al_2021, o2022modeldegradation}). Further research has considered the efficient computational implementation of the \gls{pet} model (e.g., \cite{berliner2021petlion} and citations therein) and on the lack of identifiability of some of the model parameters (e.g., \cite{berliner2021nonlinear, galuppini2023nonlinear}, and citations therein). The nominal \gls{pet} model describes internal battery phenomena, without considering the uncertainties inherent in the cell-to-cell variation of lithium-ion batteries or variations in the battery environment. Uncertainty propagation should be considered when developing fast-charging protocols, however, to ensure that battery degradation remains limited when battery operation deviates from the nominal battery model.

Most studies on \gls{pet} models ignore the effects of uncertainties on their predictions. A potential partial reason for this situation is because \gls{uq} that takes nonlinear dynamics into account via the \gls{mc} method requires on the order of ten thousand simulations for a single case study, so the computational time needed for the uncertainty analysis can be a bottleneck. The alternative approach of polynomial chaos theory evaluates uncertainty using an orthogonal polynomial expansion of the probability distribution of each random variable. Due to its significantly faster convergence speed and lower computational cost compared to the \gls{mc} method, its application to battery applications has recently become of interest including applications for the estimation of \gls{soh} and \gls{soc} \cite{bashash2013battery}, parameter sensitivity analysis \cite{streb_param_estimation2022, duong2017uncertainty, orcioni2019stochastic, hadigol2015uncertainty}, parameter estimation \cite{Fogelquist_2023, streb_reparam_2023}, and model predictive control \cite{pozzi2022stochastic}.

Here, we consider the effects of uncertainty for the fast charging of batteries, and how this information can be used to design protocols that explicitly take probabilistic uncertainties into account. We use non-intrusive \gls{pce} to quantify the effects of uncertainty, identify confidential intervals (CIs) for the key states during charging, and compare the results to the nominal case to determine the impact of uncertainty. 

This article is organized as follows. The next sections summarize the \gls{pet} model and polynomial chaos theory. Next, we present a case study of the application of polynomial chaos theory to the \gls{pet} model for the \gls{cccv} protocol, which is known to be nearly optimal for fast charging for this lithium-ion battery for operations restricted to satisfy the pre-specified current and voltage constraints \cite{matschek2023necessary}. We explicitly consider the effects of changing the time of the switching between the \gls{cc} phase and the \gls{cv} phase in response to the change in the ambient temperature and \gls{pet} model parameters. That is, we take into account C-rate and charging constraints such that the probability of violating the degradation constraints is below a pre-specified value. The article concludes with a discussion of limitations and conclusions.

\section{PET Modeling} \label{pet}

\begin{figure*}[!tb]
\centering \includegraphics[width = 0.9\linewidth]{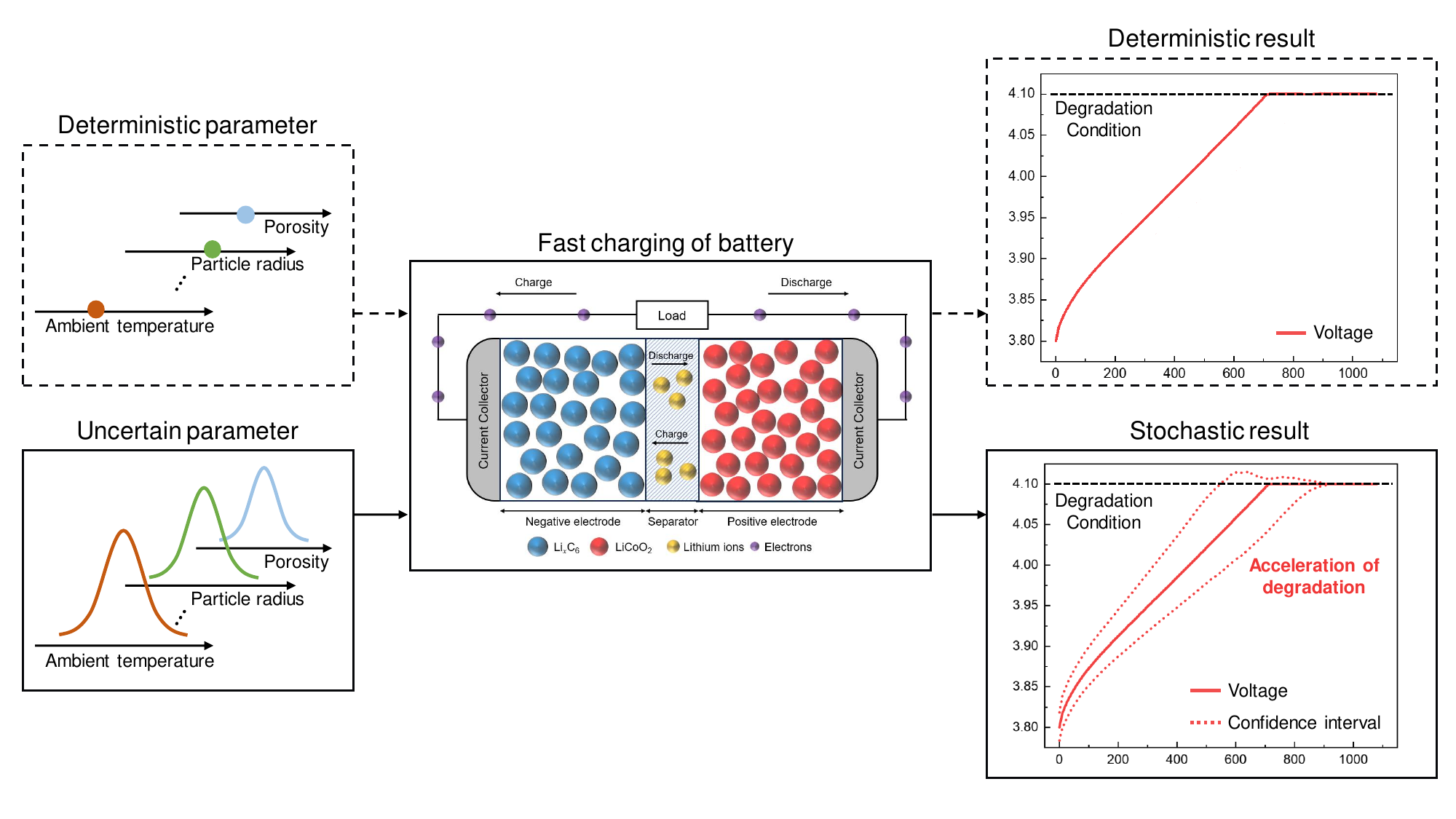}

    \vspace{-0.6cm}

\caption{Schematic of the uncertainty propagation during fast charging.}
    \label{fig:diagramPET}
\end{figure*}

The \Gls{pet} modeling framework for \glspl{lib} is a physics-based framework for modeling the cycling behavior of a cell. For \glspl{lib}, the model considers three regions: the porous anode, the electrolyte, and the porous cathode (Fig.\  \ref{fig:diagramPET}). Diffusive transport is modeled by Fick's law and electrical resistance modeled by Ohm's law. The reaction at the interface between the liquid electrolyte and the solid particles is modeled by electrochemical interfacial reaction kinetics, usually by the Butler--Volmer equation. The resulting \gls{pet} model is described by coupled partial differential-algebraic equations (DAEs). The \gls{pet} model is also referred to as the \gls{p2d} model, where the $x$ dimension is the distance across the length of the cell and the $r$ dimension is the so-called pseudo second dimension which describes diffusion inside porous electrode particles. More details of the model can be found in the original publication by Newman and coworkers \cite{NEWMANN1975PET}. 

When a load is connected to a charged \gls{lib}, some electrons leave the anode, powering the load, and some electrons enter the cathode. Inside the cell at the interface of porous graphite particle (anode, i.e., negative electrode), lithium ions enter the liquid electrolyte, leaving electrons behind, travel through the electrolyte and separator to reach a porous cathode particle (positive electrode), and electrochemically react again forming a lithium-metal-oxide, e.g., LiCoO$_2$ (Fig.\ \ref{fig:diagramPET}). 
The complexity of the underlying reactions and potentially also parasitic reactions when degradation is considered leads to numerous model parameters, such as diffusion constants, reaction constants, and geometric electrode properties. A detailed nonlinear identifiability analysis indicates that many of the model parameters are uncertain when fit to the experimental data collected for lithium-ion batteries \cite{berliner2021nonlinear}.

\section{Polynomial Chaos Theory} \label{pce}

\gls{uq} analyzes the impact of probabilistically uncertain parameters or inputs on the model states or outputs. The \gls{mc} method is most commonly used in \gls{uq}. This method is based on statistics; when the number of samples is $n$, the convergence rate scales as $ 1/\sqrt{n}$ \cite{shen2016polynomial}. Accurately quantifying system uncertainty using the \gls{mc} method requires many model evaluations, usually on the order of ten thousand or more, and is a computationally expensive task. Polynomial chaos theory is an alternative method to quantify the statistical information of the response while being orders of magnitude faster. As such, polynomial chaos theory is an interesting alternative to deal with the uncertainties of a complex dynamical system compared to the \gls{mc} method. Polynomial chaos theory can quantify the probability distribution function for any state, which can be plotted in the form of a histogram or can be used to construct prediction intervals for a variety of confidence levels for simpler visualization. Compared to the \gls{mc} method, polynomial chaos theory has been shown to achieve similar accuracy in the quantification of the effects of probabilistic uncertainties while requiring much less computational cost \cite{xiu2002wiener}. 

Polynomial chaos theory uses orthogonal polynomials to quantify the effects of uncertainty on the system response caused by the effects of probabilistic inputs or parameters. Polynomial chaos theory can directly handle a large variety of distributions on the input parameters, including uniform and Gaussian, based on the Askey scheme \cite{askey1985some}. Each distribution has a corresponding the optimal set of orthogonal polynomials as a basis, as shown in Table \ref{tab:PCEbasis}.

\begin{table}[!tb]

\vspace{0.2cm}

\centering
\caption{\label{tab:PCEbasis} Correspondence between the type of distribution and polynomial basis}

\vspace{-0.2cm}

\resizebox{8.7cm}{!}{%
\begin{tabular}{@{}c@{}c@{}c@{}c@{}}
\hline
\ \textbf{Distribution} \ & \  \textbf{Density function} \ & \  \textbf{Orthogonal polynomial} \ & \  \textbf{Hilbertian polynomial} \ \\ \hline
Normal        &  $\frac{1}{\sqrt{2\pi }}e^{-{x^2}/{2}}$  & Hermite $H_n(x)$       &       $H_n(x)/\sqrt{k!}$                        \\
Uniform          &       ${1}/{2}$        & Legendre $P_{n}(x)$   &           $P_{n}(x)/{\sqrt{{1}{2k+1}}}$                     \\ \hline
\end{tabular}
}

\vspace{-0.3cm}
\end{table}

Assume that the computationally intensive model $\textit{M}$ has $n$ input parameters. The input vector is defined as $\mathbf{X}=\left\{ x_{(1)}, x_{(2)}, x_{(3)}, \cdots{}, x_{(n)}\right\}$ whose elements are assumed to be independent variables.

The univariate orthogonal polynomial determined according to the distribution of the input is expressed as a Hilbertian basis $\psi_{k}(x)$ through normalization \cite{dammak2019numerical, xiu2003modeling},
\begin{equation}
\label{eq1}
\Psi_{\alpha}(\textbf{X})=\prod_{i=1}^{n}\psi_{\alpha_{i},(i)}(x_{i}).
\end{equation}

Multivariate orthonormal polynomials, $\Psi_{\alpha}(\textbf{X})$, can be obtained from the tensor products of the above univariate orthonormal polynomials $\psi_{k}(x)$,
\begin{equation}
\label{eq2}
\mathit{M}(\textbf{X})=Y=\sum_{\alpha\in \mathbb{N}^{n}}\textbf{a}_{\alpha}\Psi_{\alpha}(\textbf{X}).
\end{equation}

The \gls{pce} for $Y$ is defined for the input random vector $X$ and expands to the infinite series (\ref{eq2}). In engineering applications, the series is reduced to being finite through truncation. The model reduced through the truncation follows the set $A^{n,p}=\left\{\alpha| \alpha \in \mathbb{N}^{n}, \left| \alpha \right|\le p \right\}$, with $ \left| \alpha \right|$ always less than or equal to the maximum degree $p$, 
\begin{equation}
\label{eq3}
\text{card}\,A^{n,p}\equiv P = \binom{n+p}{p}=\frac{(n+p)!}{n!p!}.
\end{equation}
The reduced polynomial expansion $Y^{PC}$ is represented by
\begin{equation}
\label{eq4}
\textit{M}^{PC}(\textbf{X})=Y^{PC}=\sum_{\alpha \in A^{n, p}}\textbf{a}_{\alpha}\Psi_{\alpha}(\textbf{X}).
\end{equation}

The coefficient of truncated \gls{pce} is calculated using a non-intrusive approach, where the least-squares minimization method was used \cite{berveiller2006stochastic}. This method identifies the coefficients that minimize the truncation error of infinite and truncated polynomial expansions,
\begin{equation}
\label{eq5}
\hat{\textbf{a}}_{\alpha}=\text{argmin}\, \text{E}\!\left[ \mathcal{M}(\textbf{X})-\mathcal{M}^{PC}(\textbf{X}) \right]\!.
\end{equation}

The uncertainty of the system is quantified using statistical moments calculated through polynomial coefficients, such as mean and variance, and each coefficient can be calculated through the orthonormality of the polynomial basis. The mean is expressed as the constant term $a_{0}$ of the polynomial expansion, and the variance is expressed as the sum of the squares of all coefficients excluding the constant term $\sum_{\alpha\in A^{n,p}}a_{\alpha}^{2}-a_{0}^{2}$.

Moreover, the calculation of variance through simple computation of coefficients allows for variance-based global sensitivity analysis. The Sobol' indices is suitable for analyzing the sensitivity of parameters in complex systems such as \gls{lib} because it does not assume that the system is linear \cite{homma1996importance, sudret2008global}. Here, the analysis is done using total indices, which also take into account interactions between parameters.

\section{Case Studies} \label{cccv}
This section considers the impact of probabilistic uncertainty propagating to the state of charge. The process of propagating from the input parameters of the \gls{pet} model begins with computing the nominal simulation. There is evidence that high temperatures, voltages, and C-rates during battery charging accelerate battery degradation. Therefore, it is reasonable to introduce thresholds to minimize degradation: temperature (e.g., $>$ 40$^\circ$C) and voltage (e.g., $>$ 4.1 V) \cite{leng2015effect, klein2011optimal}. Additionally, side reactions cause lithium plating on the electrode, irreversibly reducing battery capacity, which is considered above 0 V because the lithium-plating overpotential accelerates as it decreases below 0 V.

We demonstrate the capabilities of \gls{pce} for quantification of the effects on uncertainties on the states for a \gls{cccv} fast-charging protocol, which is known to be nearly optimal for the lithium-ion battery model considered in this study when subject to operating below maximum current and voltage constraints \cite{matschek2023necessary}.
We also consider adjusting the transition time from constant current to constant voltage charging to ensure that the probability of violating the degradation constraints is below some pre-specified value. 

\subsection{CC-CV Charging}
The effect of uncertain parameters during charging is identified through the \gls{cccv} protocol. We analyzed charging states from 20\% to 80\% \cite{faisal2021fuzzy, garg2021high}, which are frequently considered the \gls{soc} range for fast charging of electronic vehicles. In \gls{cccv} charging protocol, when the maximum voltage is reached in the \gls{cc} phase, it switches to the \gls{cv} phase and charging ends when \gls{soc} reaches 80\% or the maximum temperature and minimum lithium-plating overpotential are reached.

\begin{table*}[h]
\centering

\vspace{0.2cm}

\caption{\label{tab:InputParam} Description of Uncertain Parameters \cite{hadigol2015uncertainty}}

\vspace{-0.2cm}

\resizebox{14cm}{!}{%
\begin{tabular}{ccccc}
\hline
Parameter & Description                          & Unit             & Reference value & Random input                       \\ \hline
$T_\text{amb}$      & Ambient temperature                  & K                & 298.15        & Gaussian, $\mu=298.15$, $\sigma=1.0$       \\
$D^{s}_{p}$       & Positive solid-phase diffusivity     & m$^{2}$s$^{-1}$            & 1.0$\times$10$^{-14}$       & Uniform, {[}0.9$\times$10$^{-14}$, 1.1$\times$10$^{-14}${]}    \\
$D^{s}_{n}$       & Negative solid-phase diffusivity     & m$^{2}$s$^{-1}$            & 3.9$\times$10$^{-14}$       & Uniform, {[}3.51$\times$10$^{-14}$, 4.29$\times$10$^{-14}${]}  \\
$k_{p}$        & Positive reaction rate constant      & m$^{2.5}$mol$^{-0.5}$s$^{-1}$ & 2.334$\times$10$^{-11}$     & Uniform, {[}2.1$\times$10$^{-11}$, 2.56$\times$10$^{-11}${]}   \\
$k_{n}$        & Negative reaction rate constant      & m$^{2.5}$mol$^{-0.5}$s$^{-1}$ & 5.031$\times$10$^{-11}$     & Uniform, {[}4.52$\times$10$^{-11}$, 5.53$\times$10$^{-11}${]}  \\
$D_{p}$        & Positive electrolyte diffusivity     & m$^{2}$s$^{-1}$            & 7.5$\times$10$^{-10}$       & Uniform, {[}6.75$\times$10$^{-10}$, 8.25$\times$10$^{-10}${]}  \\
$D_{s}$        & Separator electrolyte diffusivity    & m$^{2}$s$^{-1}$            & 7.5$\times$10$^{-10}$       & Uniform, {[}6.75$\times$10$^{-10}$, 8.25$\times$10$^{-10}${]}  \\
$D_{n}$        & Negative electrolyte diffusivity     & m$^{2}$s$^{-1}$            & 7.5$\times$10$^{-10}$       & Uniform, {[}6.75$\times$10$^{-10}$, 8.25$\times$10$^{-10}${]}  \\
$L_{a}$        & Positive current collector thickness & m                & 1.0$\times$10$^{-5}$        & Uniform, {[}0.8$\times$10$^{-5}$, 1.2$\times$10$^{-5}${]}      \\
$L_{p}$        & Positive electrode thickness         & m                & 8.0$\times$10$^{-5}$        & Uniform, {[}7.7$\times$10$^{-5}$, 8.3$\times$10$^{-5}${]}      \\
$L_{s}$        & Separator collector thickness        & m                & 2.5$\times$10$^{-5}$        & Uniform, {[}2.2$\times$10$^{-5}$, 2.8$\times$10$^{-5}${]}      \\
$L_{n}$        & Negative electrode thickness         & m                & 8.8$\times$10$^{-5}$        & Uniform, {[}8.5$\times$10$^{-5}$, 9.1$\times$10$^{-5}${]}      \\
$L_{z}$        & Negative current collector thickness & m                & 1.0$\times$10$^{-5}$        & Uniform, {[}0.8$\times$10$^{-5}$, 1.2$\times$10$^{-5}${]}      \\
$\epsilon_{p}$        & Positive porosity                    & --                & 0.385         & Uniform, {[}0.36, 0.41{]}          \\
$\epsilon_{s}$        & Separator porosity                   & --                & 0.724         & Uniform, {[}0.63, 0.81{]}          \\
$\epsilon_{n}$        & Negative porosity                    & --                & 0.485         & Uniform, {[}0.46, 0.51{]}          \\
$R^{p}_{p}$       & Positive particle radius             & m                & 2.0$\times$10$^{-6}$        & Gaussian, 
$\mu=2.0$$\times$10$^{-6}$, $\sigma=0.3896$$\times$10$^{-6}$ \\
$R^p_n$       & Negative particle radius             & m   & 
2.0$\times$10$^{-6}$        & Gaussian, $\mu=2.0$$\times$10$^{-6}$, $\sigma=0.1354$$\times$10$^{-6}$ \\
$\text{Brugg}_{p}$    & Positive Bruggeman coefficient       & --                & 4.0           & Uniform, {[}3.8, 4.2{]}            \\
$\text{Brugg}_{s}$    & Separator Bruggeman coefficient      & --                & 4.0           & Uniform, {[}3.8, 4.2{]}            \\
$\text{Brugg}_{n}$    & Negative Bruggeman coefficient       & --                & 4.0           & Uniform, {[}3.8, 4.2{]}            \\
$t_{+}$        & Transference number                  & --                & 0.364         & Uniform, {[}0.345, 0.381{]}        \\
$\sigma_{p}$        & Positive electronic conductivity                  & S m$^{-1}$                & 100         & Uniform, {[}90, 110{]}        \\
$\sigma_{n}$        & Negative electronic conductivity                  & S m$^{-1}$                & 100         & Uniform, {[}90, 110{]}\\ \hline
\end{tabular}
}

\vspace{-0.4cm}
\end{table*}

\begin{figure*}[h!]

    \centering
    \includegraphics[width=0.8\textwidth]{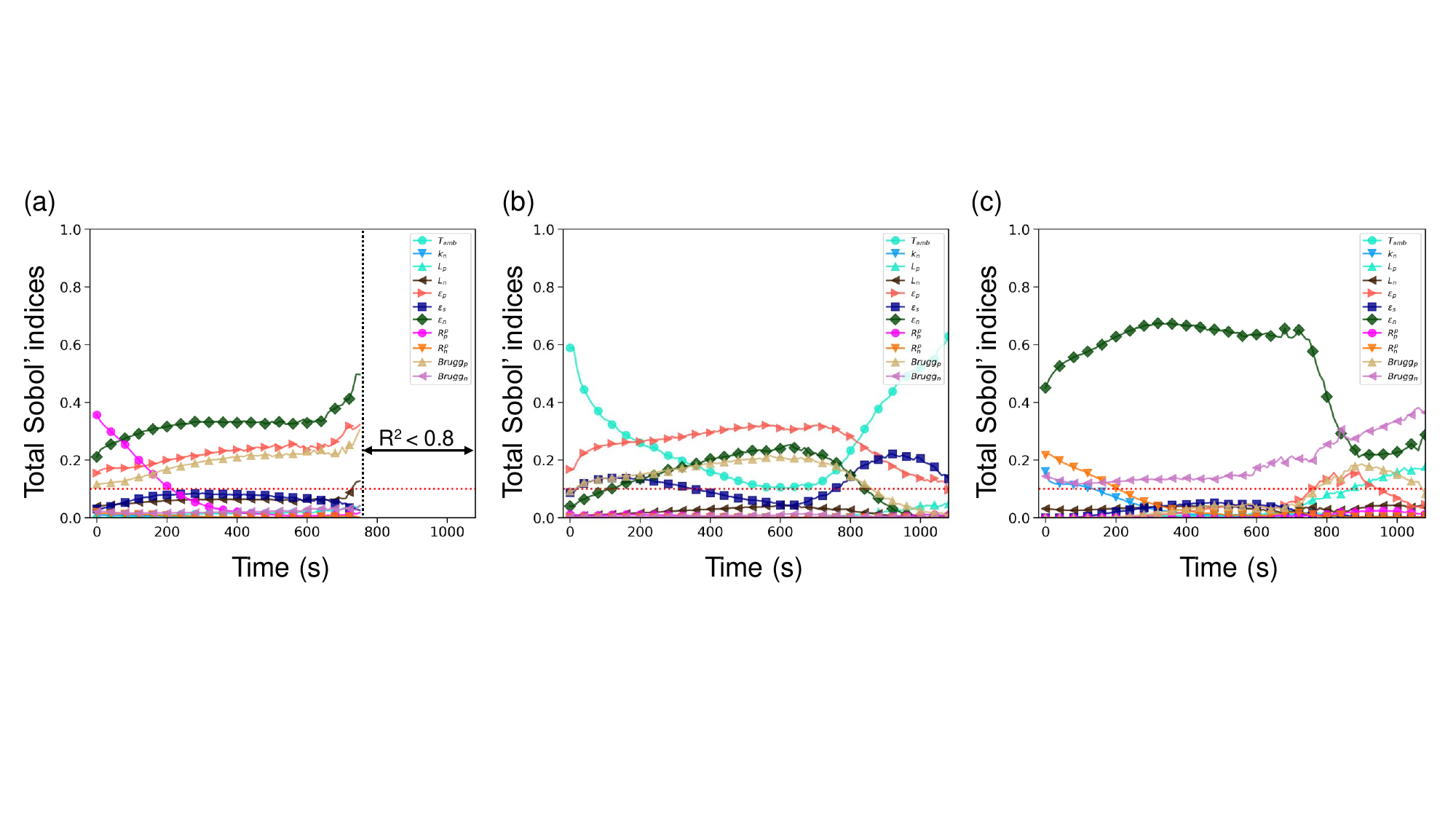} 

    \vspace{-2.5cm} 

    \caption{Total Sobol' indices for three degradation conditions for 2.2C CC-CV charging: (a) voltage, (b) temperature, (c) lithium-plating overpotential.}
\label{fig:Sobol}
\end{figure*}

Statistical information about the \gls{qoi} of the three degradation conditions during \gls{cccv} charging is obtained through \gls{pce}. If the R-squared (R$^{2}$) of the \gls{pce} is lower than 0.8, it is determined to have low reliability and is excluded from the results. 95\% \gls{ci} of \gls{qoi} analyzed at 10-second intervals for \gls{cccv} charging are applied for three degradation conditions. Based on the charging protocol considering the \gls{ci}, the constraints of the charging protocol and C-rate settings are discussed.

Figure \ref{fig:Sobol} shows the total Sobol' indices for each degradation condition during \gls{cccv} charging. For voltage, voltage control in \gls{cv} mode reduces R$^{2}$ below 0.8. In that range, Sobol' indices and \gls{ci} are not considered due to the low reliability of \gls{pce}. Among the 24 uncertain parameters specified in Table \ref{tab:InputParam}, 11 high sensitivity parameters (Sobol’ indices $>$ 0.1) for the three degradation conditions are identified as
\begin{itemize}
\item Voltage: $L_{n}$, $\epsilon_{p}$, $\epsilon_{n}$, $R^{p}_{p}$, $\text{Brugg}_{p}$
\item Temperature: $T_\text{amb}$, $\epsilon_{p}$, $\epsilon_{s}$, $\epsilon_{n}$, $\text{Brugg}_{p}$
\item Lithium-plating overpotential: $k_{n}$, $L_{p}$, $\epsilon_{p}$, $\epsilon_{n}$, $R^{p}_{n}$, $\text{Brugg}_{p}$, $\text{Brugg}_{n}$
\end{itemize}

\begin{figure*}[h!]
    \centering
    \includegraphics[trim={0.1cm 0 0.1cm 0},clip, width =0.8\textwidth]{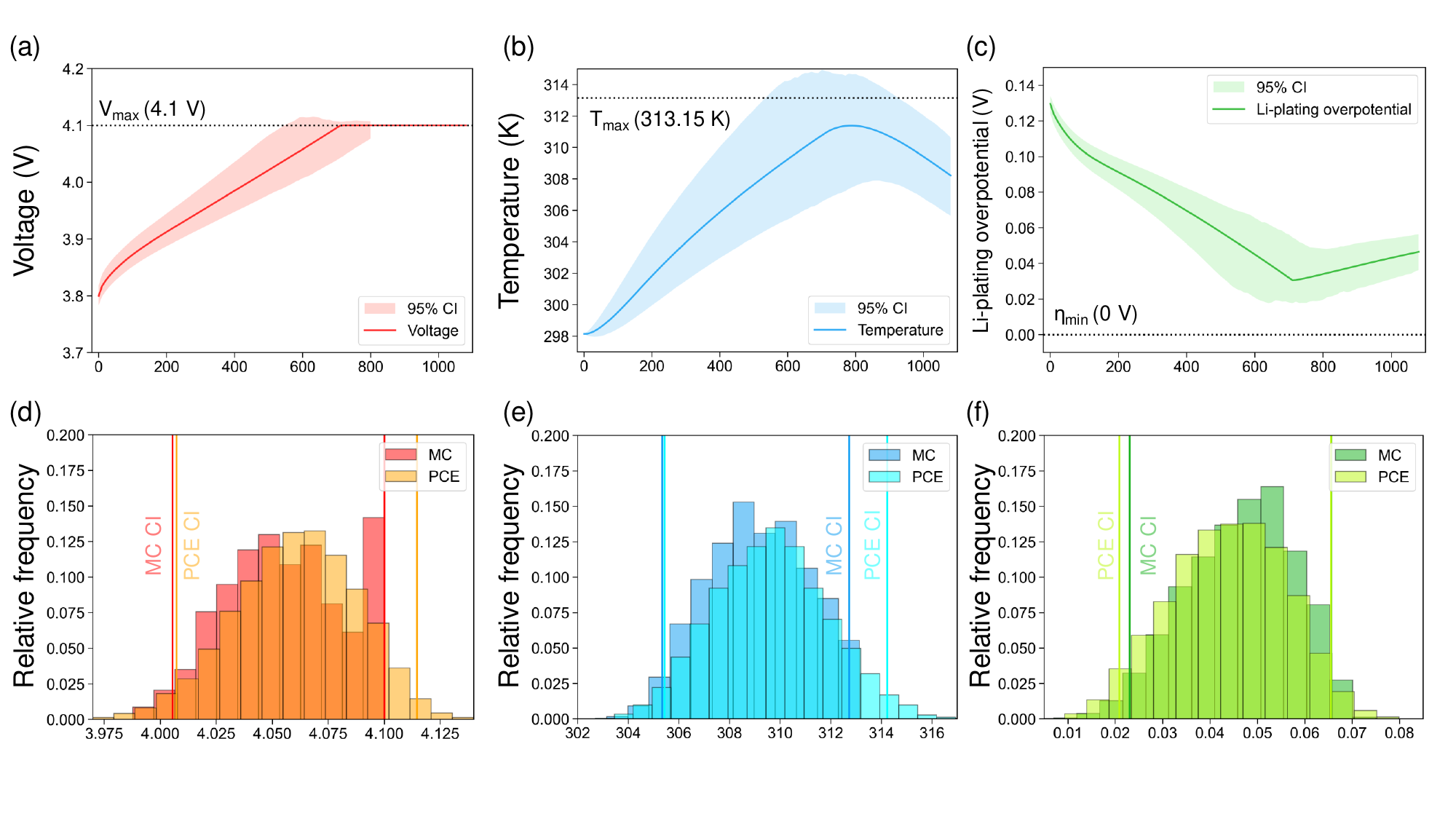}

\vspace{-0.9cm}

    \caption{Nominal values and CI for three degradation conditions in 2.2C CC-CV: (a) voltage, (b) temperature, (c) lithium-plating overpotential. QoI comparison by MC and PCE at 600 seconds: (d) voltage, (e) temperature, (f) lithium-plating overpotential.}
\label{fig:CCCV_2.2C}
\end{figure*}

The union of the sets determined with highly sensitive parameters for each condition was identified as the key parameter set for accelerated degradation.

Figure \ref{fig:CCCV_2.2C}abc shows the nominal results and \gls{ci} of voltage, temperature, and lithium-plating overpotential for 2.2C \gls{cccv} charging of LiC$_{6}$/LiCoO$_{2}$ depicted through the parameters of Ref.\ \cite{lionsimba}. Non-intrusive \gls{pce}s are generated using 300 samples. In addition, the \gls{qoi} approximated by \gls{pce} for the three degradation conditions near the time of switching to the \gls{cv} stage is evaluated at 10,000 samples and compared to \gls{mc} using 3,000 samples (Fig.\ \ref{fig:CCCV_2.2C}def). According to Table \ref{tab:Time}, \gls{pce} generated through key parameters takes only about 10.1\% of the computational budget compared to \gls{mc}.

\begin{figure*}[h!]
    \centering
    \includegraphics[trim={0.1cm 0 0.1cm 0},clip, width =0.84\textwidth]{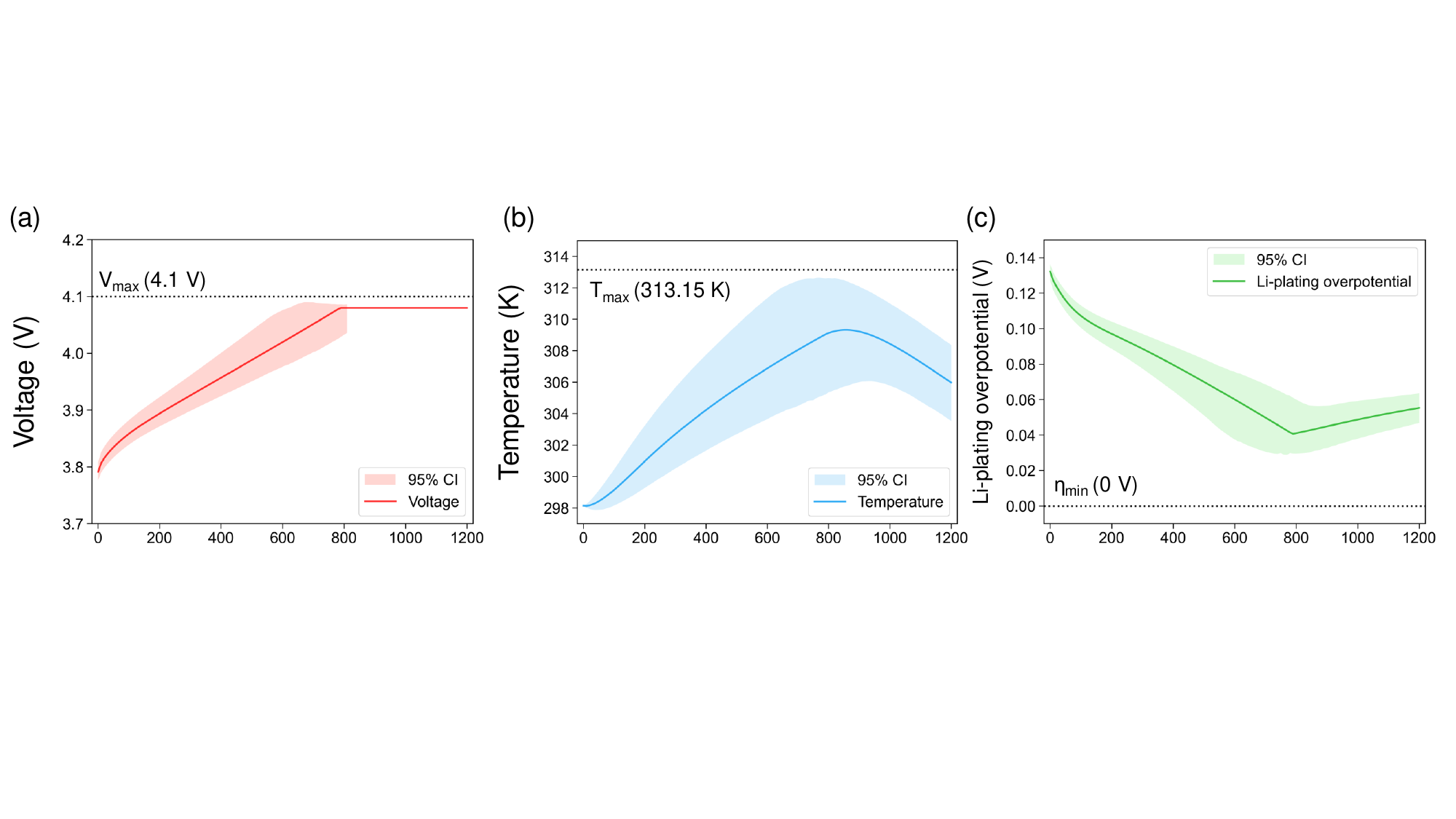}

\vspace{-3cm}

    \caption{Nominal values and CI for three degradation conditions in 2.0C CC-CV (V$_{\max}$ = 4.08 V): (a) voltage, (b) temperature, (c) Li-plating overpotential.}
\label{fig:CCCV_2C}
\end{figure*}

\begin{table}[h!]
\centering

\vspace{0.2cm}

\caption{\label{tab:Time} Computational times in CC-CV applications between Monte Carlo, PCE using 24 parameters, and PCE using 11 parameters at 600 seconds}

\vspace{-0.2cm}

\begin{tabular}{ccc}
\hline
          & MC (24 parameters)    & PCE (11 parameters) \\ \hline
 Time (s) & 3,997 & 403 \\ \hline
\end{tabular}

\vspace{-0.4cm}

\end{table}

The nominal model indicates that the charging should be switched to \gls{cv} mode in about 711 seconds when $V$ reaches $V_{\max}$ (Fig.\ \ref{fig:CCCV_2.2C}a). The upper bound of the voltage \gls{ci} reaches $V_{\max}$ at about 550 seconds. As such, in battery modules or packs composed of many cells, voltage constraints may be reached locally due to cell variability, which may lead to uneven degradation. Skewness is observed in the distribution as the \gls{cv} mode is approached, which resullts in the \gls{ci} by \gls{pce} overestimating the $V_{\max}$ (Fig.\ \ref{fig:CCCV_2.2C}d).
The nominal temperature rises to 311.4 K during charging, which does not reach the accelerated degradation condition, but the \gls{ci} exceeds the $T_{\max}$ of 313.15 K from about 530 to 930 seconds (Fig.\ \ref{fig:CCCV_2.2C}b). The voltage switches to the \gls{cv} phase when it reaches $V_{\max}$, but charging is terminated when the temperature reaches $T_{\max}$ to minimize performance degradation such as irreversible decomposition of the electrolyte. For this reason, a truncated distribution rather than skewness is observed for temperature, and the lowered R$^{2}$ is the cause of the upper bound of \gls{ci} predicting a value of temperature higher than $T_{\max}$ (Fig.\  \ref{fig:CCCV_2.2C}e). Degradation does not occur due to the $T_{\max}$ constraint, but the desired \gls{soc} (i.e., 80\%) is not reached as charging termination. In other words, uncertainty propagation in 2.2C \gls{cccv} charging indicates that degradation is accelerated by voltage, or charging is prematurely terminated by reaching temperature constraints. Since the lithium-plating overpotential does not reach the constraint during charging, the distributions identified by \gls{mc} and \gls{pce} during charging are quite similar (Fig.\  \ref{fig:CCCV_2.2C}f).

\begin{table}[]
\centering

\vspace{0.2cm}

\caption{\label{tab:Time_CCCV_comparison} Comparison of charging time and accelerated degradation according to C-rate and V$_{\max}$ for CC-CV charging}

\vspace{-0.2cm}

\begin{tabular}{ccc}
\hline
C-rate       & 2.2C  & 2.0C  \\
V$_{\max}$ (V)                & 4.1         & 4.08        \\ \hline
CC to CV (s)            & 711.2       & 787.3       \\
Total charging time (s) & 1086.2      & 1204.8      \\
Degradation             & accelerated & accelerated \\ \hline
\end{tabular}

\vspace{-0.3cm}
\end{table}

\begin{figure}[h!]
\hspace{-.15in}
    \includegraphics[trim={3cm 0cm 1.2cm 1cm},clip,width = 0.55\textwidth]{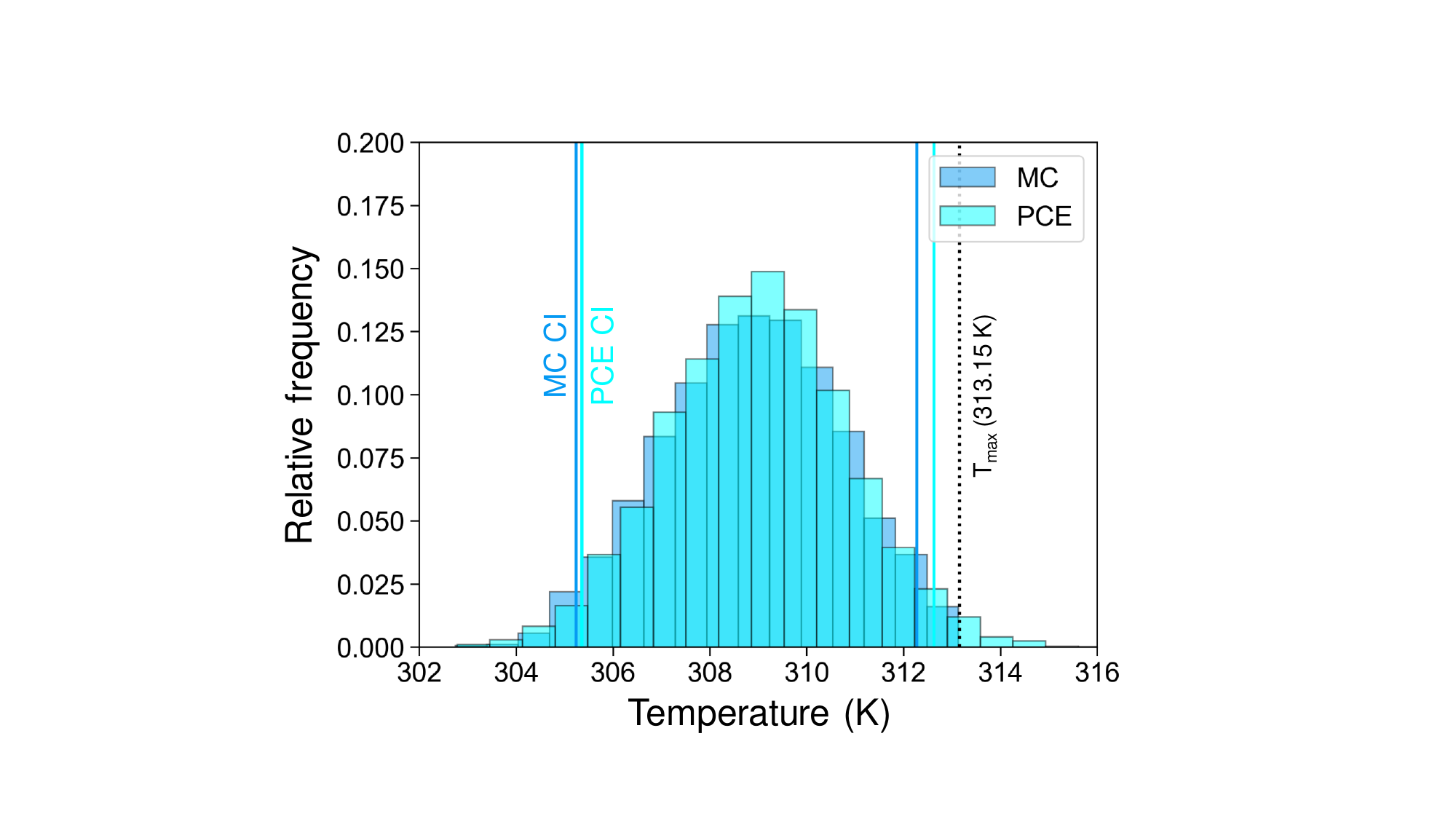}

    \vspace{-0.9cm}

    \caption{Probability distribution and \gls{ci} by \gls{mc} and \gls{pce} at 800 seconds in 2.0C \gls{cccv} charging.}
    \label{fig:CCCV_hist}
\end{figure}

Figure \ref{fig:CCCV_2C}abc shows that the degradation condition is not reached for C-rate of 2C. As $V_{\max}$ decreases to 4.09 V, the upper bound of the \gls{ci} does not reach 4.1 V. Reducing the C-rate to 2C also reduces the temperature rise so that the temperature \gls{ci} does not reach $T_{\max}$. Unlike charging at 2.2C, the upper bound of temperature \gls{ci} for a C-rate of 2C does not show a significant difference compared to \gls{ci} by \gls{mc} (Fig.\ \ref{fig:CCCV_hist}). The C-rate-dependent lithium-plating overpotential likewise does not reach the constraint (Fig.\  \ref{fig:CCCV_2C}c). However, charging under moderate conditions increases the charging time by about 118 seconds (Table \ref{tab:Time_CCCV_comparison}).

\section{Conclusion and Future Direction} \label{conclusion}

This study analyzes the impact of uncertainty on battery degradation during charging using the \gls{pet} model which expresses various internal phenomena of a battery. The \gls{pet} model has been shown to be effective in modeling \gls{lib}s, but its nominal predictions can be sensitive uncertainties in environmental and model parameters. \gls{pce} is applied to \gls{cccv} charging to identify parameters sensitive to fast charging conditions and investigate their effect. Among the 24 parameters consisting of ambient temperature and 23 \gls{pet} model parameters, only 11 parameters were identified to affect the charging status. In nominal results, once the maximum voltage is reached in the \gls{cc} stage, the degradation is not accelerated due to the transition to the \gls{cv} stage. On the other hand, when temperature and lithium-plating overpotential reach constraints during charging, charging is terminated. Uncertainties propagated to the degradation conditions indicate that accelerated degradation and premature charge termination may occur, which were not observed in the nominal results. Stochastically accelerated degradation can be minimized through adjustment of charging constraints such as $V_{\max}$ or C-rate. Our results indicate that uncertainty during charging should be considered to minimize battery performance degradation. This non-intrusive \gls{pce}-based approach can extract statistical information of \gls{qoi} with a significantly lower computational budget than \gls{mc} and can be successfully applied to further improved and complex state-dependent charging protocols.

\bibliography{references.bib}
\bibliographystyle{IEEEtran}

\end{document}